\documentclass[cits]{PoS}

\usepackage[latin1]{inputenc}                    
\usepackage{graphicx}                            
\usepackage{latexsym}                            
\usepackage{amsfonts}                            
\usepackage{amssymb}                             
\usepackage{amsmath}                             
\usepackage[mathscr]{eucal}                      
\usepackage{dcolumn}                             

\PoS{PoS(LAT2005)054}

\title{Diquark properties from lattice QCD }

\ShortTitle{Diquark properties from lattice QCD }

\author{\speaker{Konstantinos Orginos}\thanks{Present address:  Dept.  of Physics,
P.O. Box 8795,
College of William and Mary,
Williamsburg, VA 23185,
USA}\\
        CTP/LNS MIT\\
        E-mail: \email{orginos@lns.mit.edu}}

\abstract{  It has been argued recently that diquarks, a pair of
  quarks in the anti-triplet representation of SU(3) color, are
  important building blocks of baryons. The assumption that the scalar
  diquark is tightly bound seems to be nicely accommodated by
  experimental data. In this paper I attempt to extract
  phenomenological properties of diquarks from lattice QCD
  calculations. In particular, I use the MILC 2+1 dynamical fermion lattices
  with domain wall fermion valence quarks to probe diquarks very close
  to the chiral limit. }

\FullConference{XXIIIrd International Symposium on Lattice Field Theory\\
		 25-30 July 2005\\
		 Trinity College, Dublin, Ireland}

\begin{document}

\section{Introduction}

The possibility of the existence~\cite{Diakonov:1997mm} of  narrow width
five quark hadronic states has excited a lot of theoretical and
experimental activity with an aim at finding them and explaining their
structure. A recent review of the experimental status of 
pentaquarks can be found in~\cite{Danilov:2005kt}. One model for 
 the existence of such states has been put forward by
Jaffe and Wilczek~\cite{Jaffe:2003sg}. A central feature of this model
is the existence of a tightly bound diquark.  

The diquark, as its name signifies, is an object made out of two
quarks.  The color ${\bf 3}\times{\bf 3}$ is reduced to the symmetric
${\bf 6}$ and the anti-symmetric anti-triplet $\bar{\bf 3}$
representations. It is not difficult to argue that the color
anti-triplet representation is energetically more favorable that the
symmetric one. Both one gluon exchange and the 'tHooft interactions are
attractive in this channel. In what follows the term diquark is used to
describe the color anti-triplet representation.

Simple phenomenological observations ranging from deep inelastic
scattering to the QCD spectrum features indicate that not all diquarks
are created equal. It has been argued that the anti-symmetric in
flavor scalar diquark
\begin{equation}
 q_f^a { C\gamma_5 } q_{f'}^b \epsilon_{cab} \epsilon^{ff'}
 \end{equation}
is energetically more favorable than the spin 1 diquark
 \begin{equation}
 q_f^a { C\gamma_\mu } q_{f'}^b \epsilon_{cab} \;.
 \end{equation}
In the above $f$ is a flavor index, while $abc$ are color indices. The
spin indices are suppressed.
A model based on this assumption~\cite{Wilczek:2004im} seems to 
accommodate the mass hierarchies of the observed hadron spectrum.

Although the diquark is a seemingly simple object, it is difficult to
ask questions about its properties outside the scope of a model.  For
that reason a lattice QCD calculation may be able to provide more
information about its structure than real experimental data. In this
paper I describe a calculation in Lattice QCD of the difference in
binding energy between the vector and the scalar diquark. Two more
papers in these proceedings discuss ways to address similar questions
in the context of numerical calculations~\cite{Lucini05,Faccioli05}.

\section{The method}

 Measuring a mass  for the diquark is not possible since it is a color
non-singlet object that can only exist in a bound colorless state. Together
with another quark it forms a baryon. So the mass difference of a baryon
with the vector diquark (ex. $\Delta$) from a baryon with the scalar diquark
(ex. proton) can provide information on the binding energy splitting of these
two diquarks. Unfortunately, spin dependent interactions of the quark and the
diquark are also different for the two different spin configurations.
Hence mass splittings of real baryons are not clean probes of the binding
energy difference of diquarks. On the other hand, if the mass of the third
quark is infinite, the spin dependent interactions drop out, allowing us
to probe the binding energies of the diquarks.

 We can easily construct baryons made out of an infinitely heavy quark
 and a diquark, on the lattice. The correlation function of such an
 object is
\begin{equation}
G_{\Gamma}(x,t;x,0) = \langle  u^a(x,t) { \Gamma } d^b(x,t) \epsilon_{cab}
                     P^{cc'}(x,t;x,0) 
                  \bar{d}^{a'}(x,0) { \Gamma } \bar{u}^{b'}(x,0) \epsilon_{c'a'b'}
         \rangle \;,
\end{equation}
where $u$ and $d$ are the light quark flavors and $ P^{cc'}$ is the
Wilson line connecting the source and the sink which are separated by
time $t$.  The spin matrix $\Gamma$ is either $C\gamma_5$ or
$C\gamma_\mu$ for the scalar and vector diquarks, respectively. The
goal here is to calculate the lowest mass associated with the
correlation functions $G_{\Gamma}(x,t;x,0) $ and compute the mass
difference between the mass of the state with the scalar diquark
($\Lambda_Q$) and that of the mass of the sate with the vector diquark
($\Sigma_Q$). This mass difference is the difference in the binding
energy of the two diquarks.

\section{Lattice details}
\begin{figure}
  \begin{minipage}{210pt}
    \begin{center}
    \includegraphics[width=210pt]{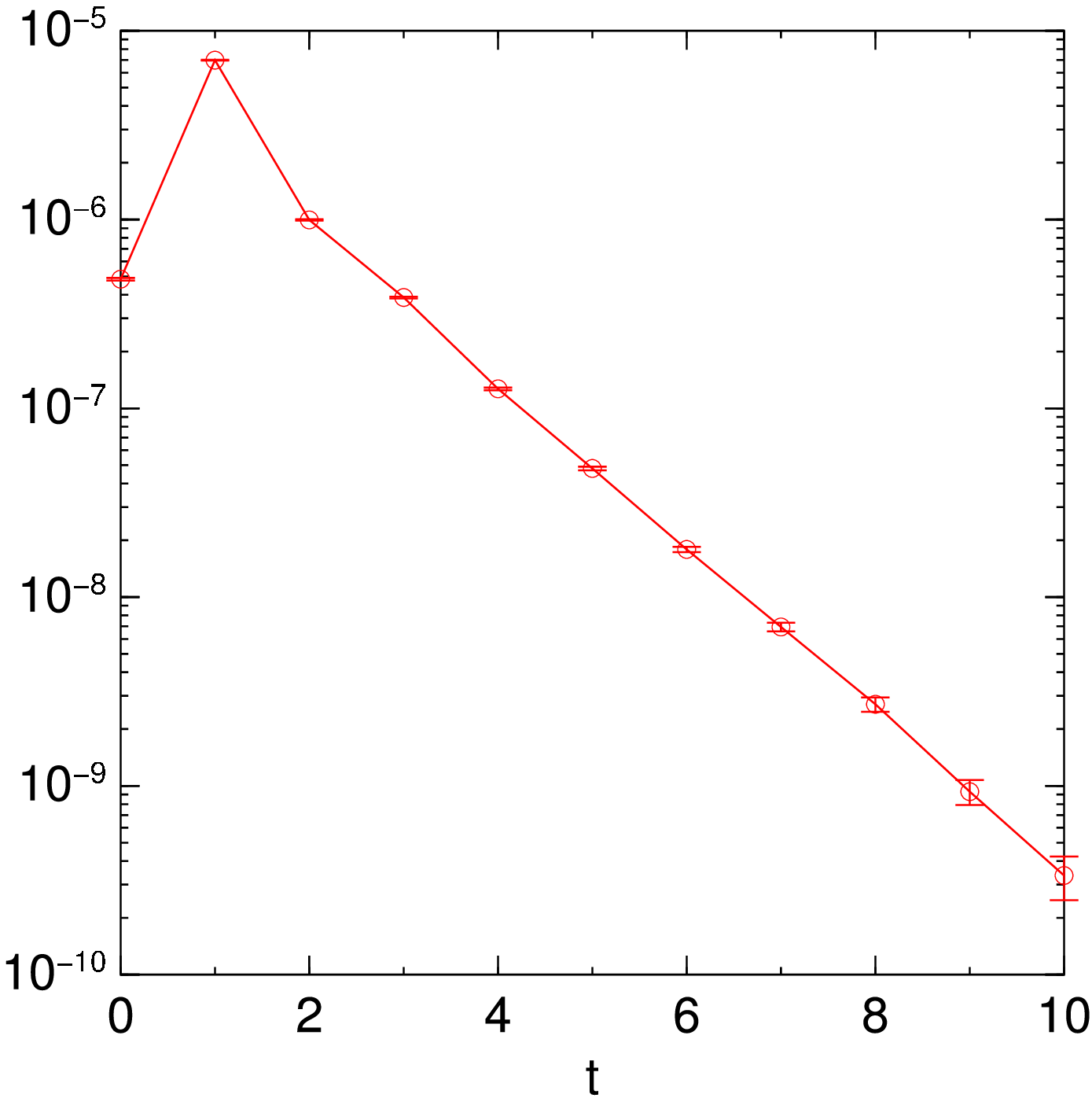}      
    \end{center}
  \end{minipage}\hspace{5pt}
  \begin{minipage}{210pt}
    \begin{center}
    \includegraphics[width=210pt]{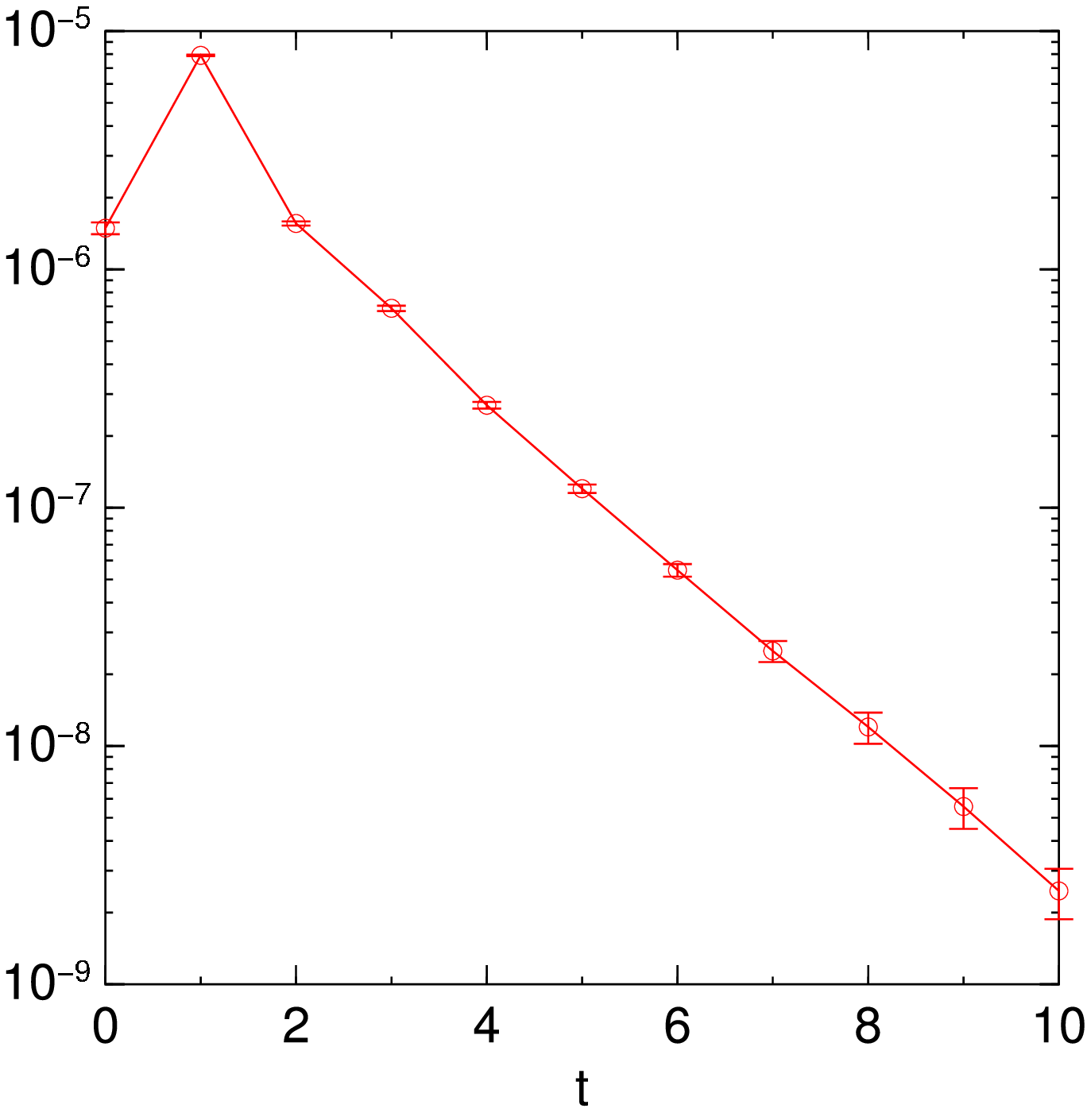}    
    \end{center}
  \end{minipage}
   \caption{\label{fig:LambdaSigma}The  $\Sigma_Q$ (left) and $\Lambda_Q$ (right) correlators for the staggered bare quark mass 0.010}
\end{figure}

In order to perform the above calculation it is essential to use quark
masses as light as possible. In addition, in order be able to obtain
physically interesting results we performed the calculation with
dynamical fermions. We used the 2+1 improved Kogut-Susskind
(Asqtad)~\cite{Orginos:1999kg,Orginos:1999cr,Toussaint:1998sa} fermion
lattices at lattice spacing $a=0.125$fm provided by
MILC~\cite{Bernard:2001av,Aubin:2004wf}. The improved Kogut-Susskind
action has been shown to have very good scaling
properties~\cite{Bernard:2000gd,Bernard:1999xx}. For the light quark
propagators we use gauge invariant Gaussian smeared source propagators
provided by LHPC. For details on the propagator generation
see~\cite{Renner:2004ck,Edwards05}. The essential feature of these
propagators is that the domain wall fermion mass has been
tuned so that the pion mass matches the Kogut-Susskind Goldstone pion
mass. In addition, HYP smearing has been used in order to improve the
domain wall fermion explicit chiral symmetry breaking. Tests of the
locality and the chiral behavior of the domain wall action have been
performed and proved that the action is local and that the explicit
chiral symmetry breaking is negligible as far as the relevant physical
observables are concerned.

\begin{figure}
\begin{minipage}{210pt}
    \begin{center}
    \includegraphics[width=210pt]{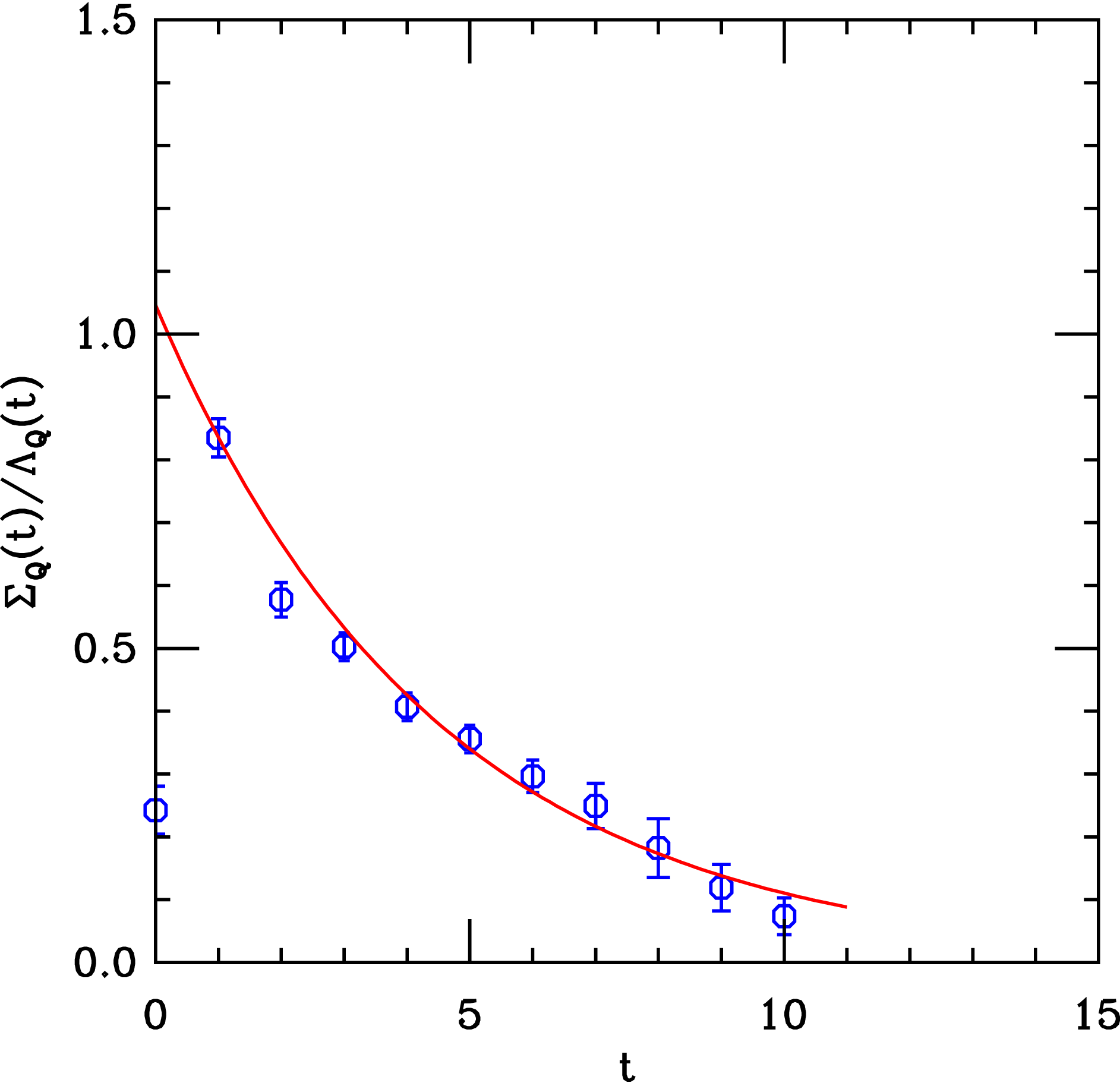}      
    \end{center}
  \end{minipage}\hspace{5pt}
  \begin{minipage}{210pt}
    \begin{center}
    \includegraphics[width=210pt]{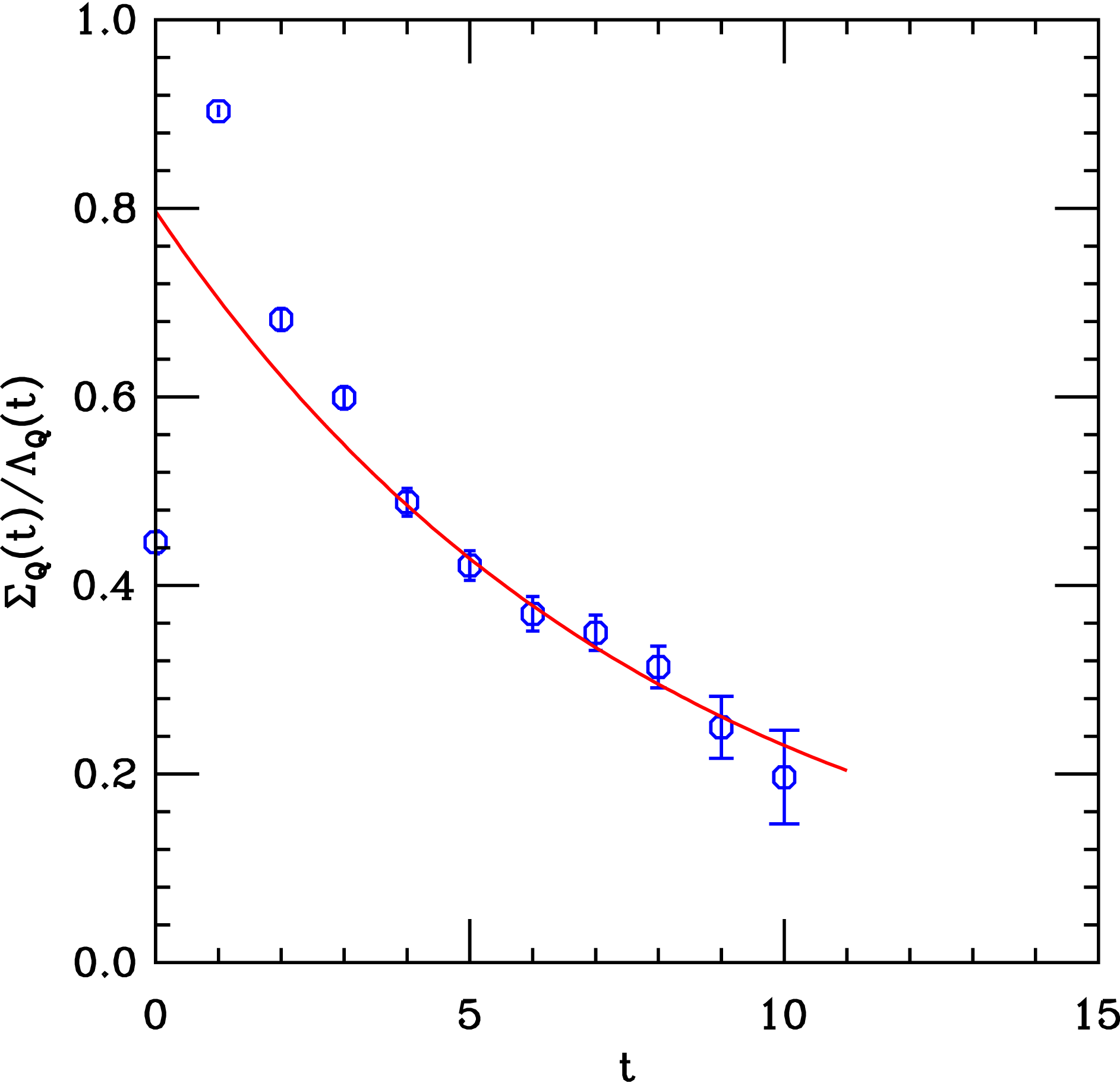}
    \end{center}
  \end{minipage}
\begin{minipage}{210pt}
    \begin{center}
    \includegraphics[width=210pt]{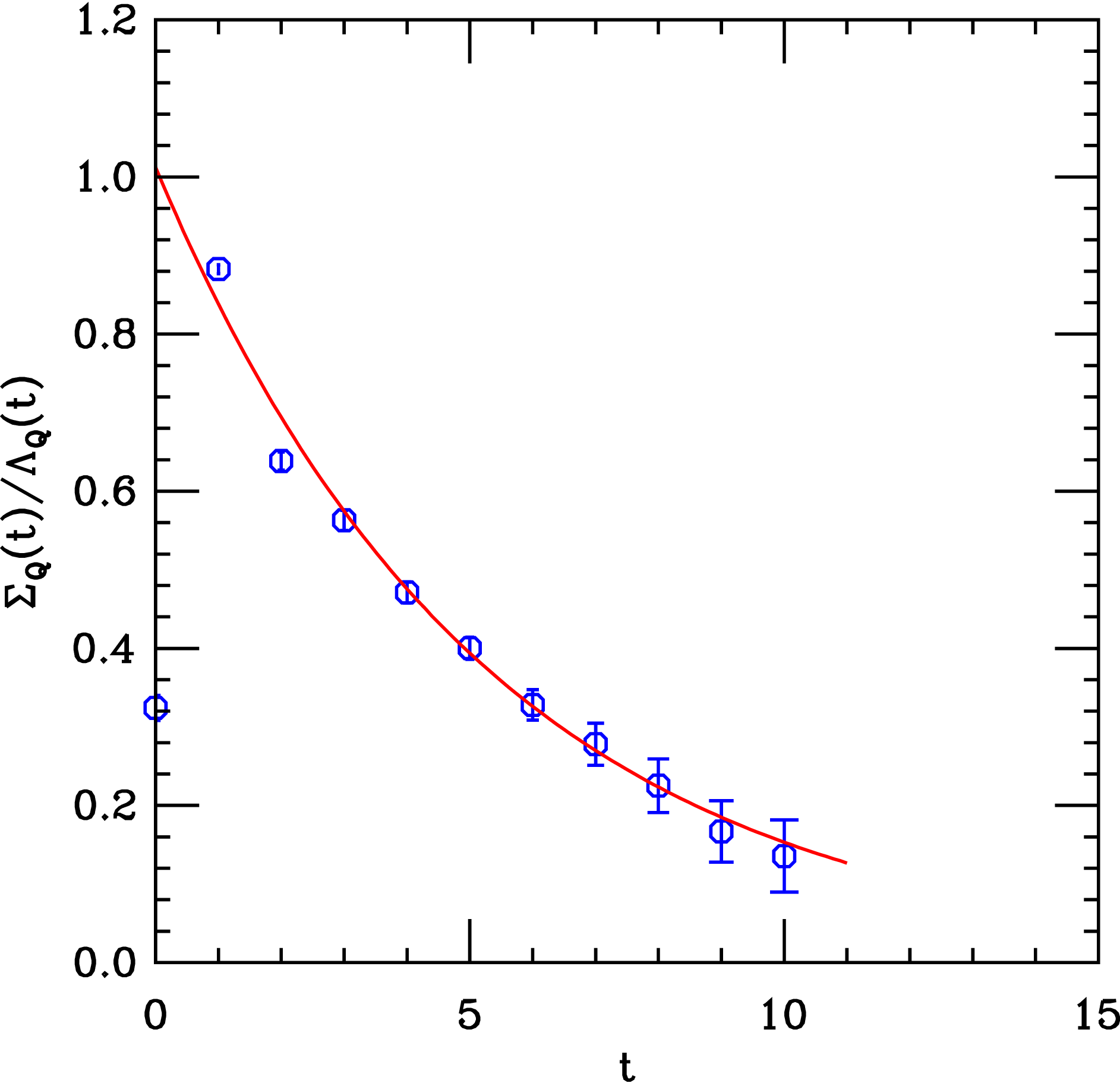}
    \end{center}
  \end{minipage}\hspace{5pt}
  \begin{minipage}{210pt}
    \begin{center}
    \includegraphics[width=210pt]{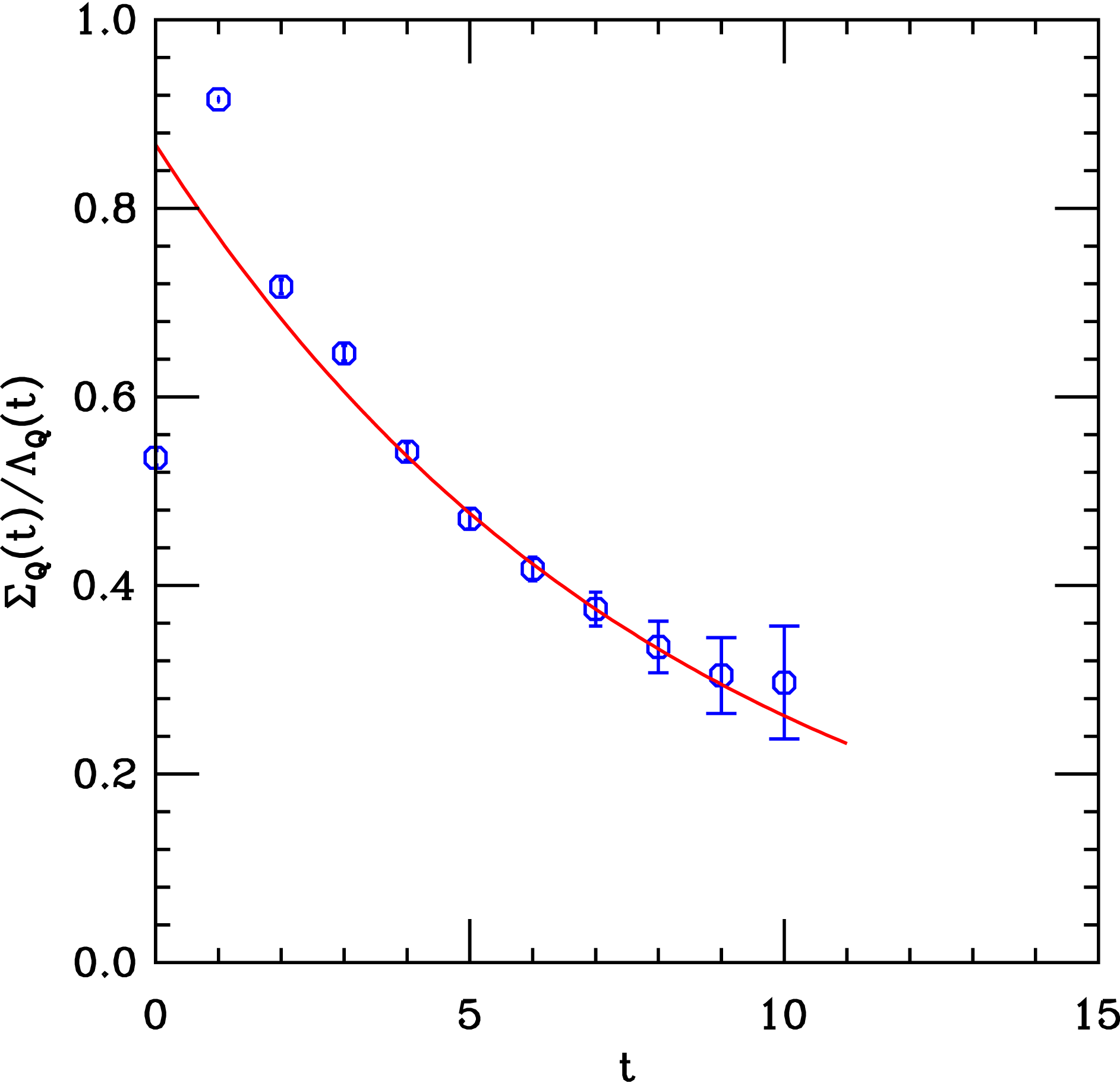}    
    \end{center}
  \end{minipage}     
   \caption{\label{fig:Ratio}The ratio $R(t)$ for the staggered bare quark masses 0.007 (top left), 0.010 (top right), 0.020 (bottom left) and 0.030 (bottom right). The red lines are exponential fits in the time range 4 to 10.}
\end{figure}

The number of configurations used range from 400 to 650 configurations
depending on the ensemble. The calculation was performed for the bare
light Kogut-Susskind quark masses 0.007, 0.010, 0.020 and 0.030.
Since we are interested in the mass difference between the $\Lambda_Q$
and $\Sigma_Q$ states, we compute the ratio of the these two
correlators
\begin{equation}
  R(t) = \frac{G_{C\gamma_\mu}(t)}{G_{C\gamma_5}(t)} \;.
\label{eq:ratio}
\end{equation}
Then we fit this ratio to a simple exponential from which we extract
the mass difference of the two low lying states. Jackknife analysis is
used to perform the error estimates. For the scale, we used the
$a=0.125$fm. For all the fits, we chose the  time range 4 to 10.
The signal for time larger than 10 deteriorated rapidly. In Figure~\ref{fig:LambdaSigma} we present a typical set of $\Lambda_Q$ and $\Sigma_Q$ correlators
for the bare staggered sea quark mass 0.010. The ratios of correlators from which we extracted the diquark mass splittings are presented in Figure~\ref{fig:Ratio}.

\section{Results and discussion}

\begin{figure}
   \begin{center}
    \includegraphics[width=250pt]{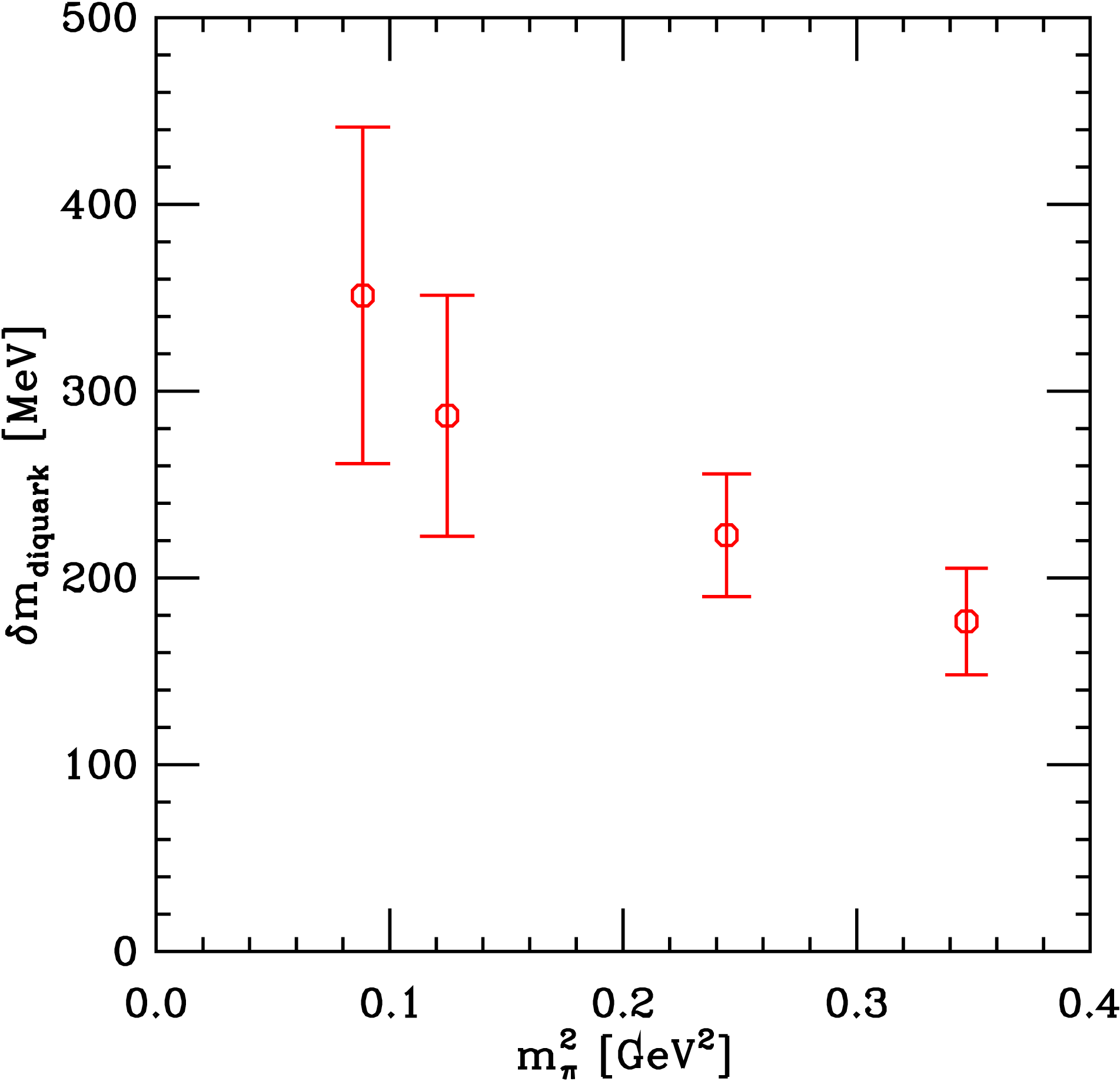}    
    \end{center}
  \caption{\label{fig:split}The diquark mass splitting in units as function of the pion mass squared.}
\end{figure}

As we can see in Figure~\ref{fig:split} the computed binding energy
difference of the scalar and the vector diquark is a rather large
number compared to QCD scales. A linear extrapolation to the
chiral limit yields a value of $360(70)$MeV. The value of the
splitting increases rapidly with decreasing quark mass providing evidence that
scalar diquarks made out of light quarks are more favorable than those
made out of heavier quarks. Note that all our quark masses are lower
than the strange quark mass.

In this calculation I have not yet addressed carefully several sources
of systematic errors such as continuum extrapolations, chiral
extrapolations, scale setting and volume dependence. Nonetheless, it
seems that all these errors should alter very little the
basic features of these results: the binding energy of the the scalar
diquark is fairly large and it increases rapidly with decreasing quark
mass.  Both these features are assumed in the Jaffe-Wilczek model of
pentaquarks~\cite{Jaffe:2003sg} and in Wilczek's picture of QCD
spectrum~\cite{Wilczek:2004im}. Here I present a numerical
justification for these assumptions.

\section*{Acknowledgments}

I thank M. Savage for helpful discussions on various aspects of this
project.  Also I thank LHPC for providing the propagators needed and
MILC for providing their 2+1 dynamical fermion lattices.  All
computations were performed on the LQCD clusters at Jlab with software
developed under SciDac (Chroma/QDP++)~\cite{Edwards:2004sx}.  This
research was partially supported by the US Department of Energy grant
DF-FC02-94ER40818.

\bibliographystyle{JHEP}
\bibliography{diquark}

\end{document}